\documentclass[prl,showpacs,twocolumn]{revtex4}
\newcommand {\etal}{{\it et al.}}
\newcommand {\TiO}{Ti$_2$O$_3$}  
\newcommand {\VO}{V$_2$O$_3$}  
\newcommand {\VOO}{VO$_2$}  
\newcommand {\TiVO}{(Ti$_{1-x}$V$_x$)$_2$O$_3$}  
\newcommand {\LSTO}{La$_{1-x}$Sr$_x$TiO$_{3+y/2}$}

\newcommand {\Neff}{$N_{\mathrm{eff}}$}

\newcommand {\dc}{$\sigma (0)$}
\newcommand {\ttwog}{$t_{2g}$}
\newcommand {\aoneg}{$a_{1g}$}
\newcommand {\aonegs}{$a_{1g}^*$}
\newcommand {\egp}{$e_g^\pi $}
\newcommand {\egps}{$e_g^{\pi *} $}

\newcommand {\EparaC}{{\bf E} $\parallel c$}
\newcommand {\EperpC}{{\bf E} $\perp c$}
\newcommand {\Tim}{$T_{\mathrm{IM}}$}

\usepackage{graphicx}
\usepackage{bm}
\usepackage{pifont}
\usepackage{amsmath}
\begin{document}
\preprint{Draft \today}
\title{Charge dynamics in thermally and doping induced insulator-metal transitions of {\TiVO}}
\author{M. Uchida$^1$, J. Fujioka$^1$, Y. Onose$^{1,2}$, and Y. Tokura$^{1,2,3}$} 
\affiliation{$^1$ Department of Applied Physics, University of Tokyo, Tokyo 113-8656, Japan \\ $^2$  Multiferroics Project, ERATO, Japan Science and Technology Agency (JST), Tokyo 113-8656, Japan \\ $^3$ Cross-Correlated Materials Research Group (CMRG), ASI, RIKEN, Wako 351-0198, Japan}
\date{}
\begin{abstract}
Charge dynamics of {\TiVO} with $x=0-0.06$ has been investigated by measurements of charge transport and optical conductivity spectra in a wide temperature range of $2-600$K with the focus on the thermally and doping induced insulator-metal transitions (IMTs). The optical conductivity peaks for the interband transitions in the 3$d$ {\ttwog} manifold are observed in the both insulating and metallic states, while their large variation (by $\sim 0.4$ eV) with change of temperature and doping level scales with that of the Ti-Ti dimer bond length, indicating the weakened singlet bond in the course of IMTs. The thermally and V-doping induced IMTs are driven with the increase in carrier density by band-crossing and hole-doping, respectively, in contrast to the canonical IMT of correlated oxides accompanied by the whole collapse of the Mott gap.
\end{abstract}
\maketitle

Insulator-metal transition (IMT) has been an important issue of condensed matter physics \cite{MIT}. In the conventional semiconductor, the IMT is achieved by the shift of the chemical potential due to the chemical or electrostatic doping, where the electronic band structure is almost rigid in the course of IMT. Even in the metallic state, the carrier density is small and proportional to the amount of doping. In typical Mott transitions of correlated $d$-electron systems such as {\VO} \cite{V2O3opt} and La$_{1-x}$Sr$_x$TiO$_3$ \cite{LSTOopt}, on the contrary, the Mott gap collapses with the temperature change or the chemical doping as a result of the reconstruction of the electronic structure on a large energy scale comparable to the Coulomb repulsion ($\sim$ several eV) upon the IMT. The carrier number estimated by the Hall coefficient is large ($\sim 1/\mathrm{site}$) compared with the dopant concentration in the metallic state. In the high-$T_{\mathrm{c}}$ cuprate system \cite{HTSC}, the parent compound is a Mott insulator (to be precise, a charge transfer insulator), but the carrier number is comparable to the dopant concentration in the low-doping and low-temperature region. One of the plausible scenarios for the origin of the doped-insulator-like feature in the normal state is the pseudogap formation due to the preformed singlet pairs or the resonating valence bond (RVB) state \cite{RVB}. 

\begin{figure}
\begin{center}
\includegraphics*[width=8.6cm]{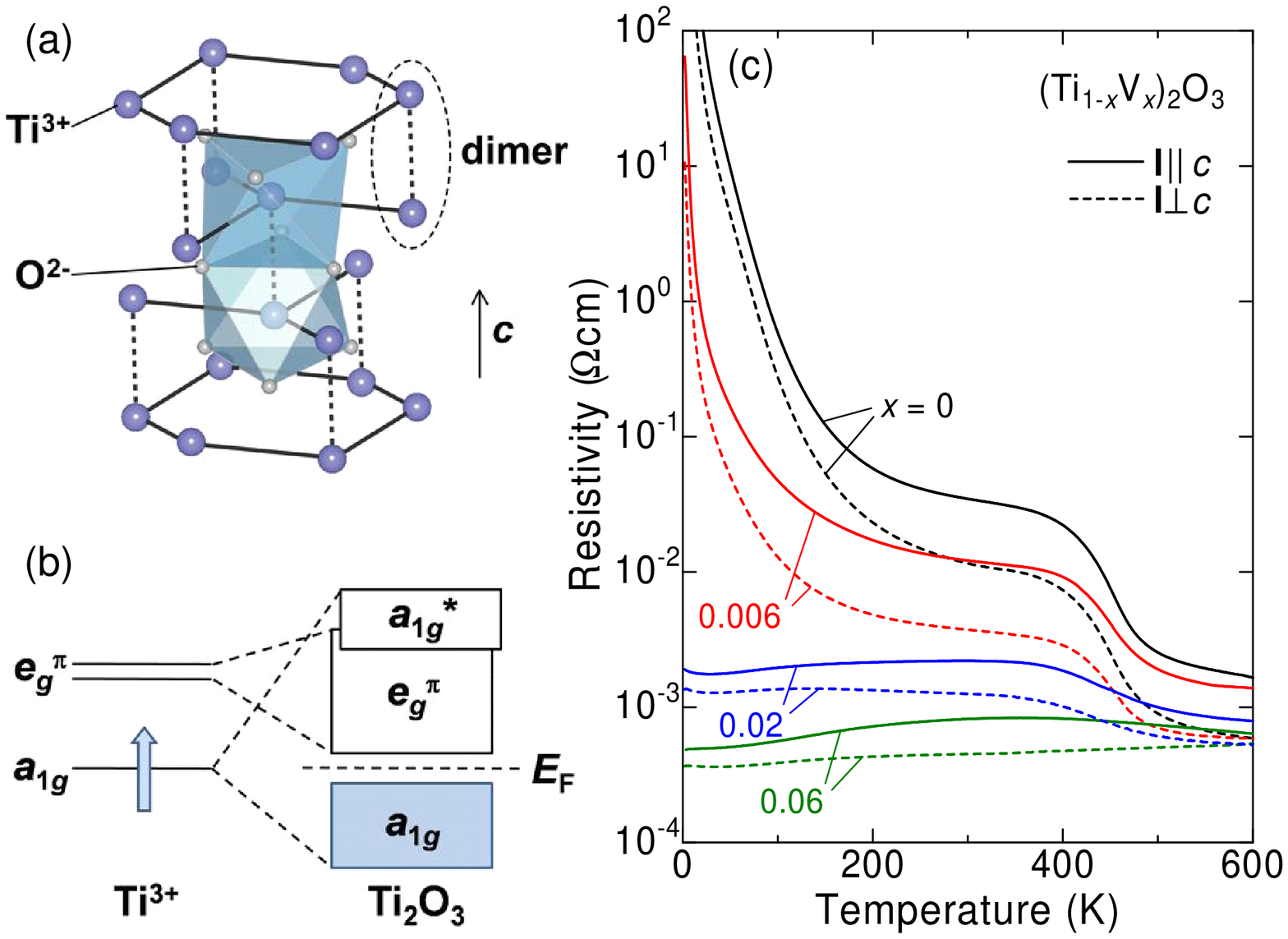}
\caption{(color online). (a) The corundum structure of {\TiO}. The dashed lines represent the bonds in Ti-Ti dimer. (b) Schematic diagram for the proposed electronic structure in {\TiO} \cite{good}. For clarity, the notation {\egp} indicates both {\egps} and {\egp} bands, as the {\egp}-{\egps} splitting is small. (c) Temperature dependence of the resistivity in {\TiVO} with $x=0$, 0.006, 0.02, and 0.06 for the current {\bf I} $\parallel c$ and {\bf I} $\perp c$.}
\label{fig1}
\end{center}
\end{figure}
{\TiO} with an $\alpha$-corundum structure (Fig. \ref{fig1}(a)) shows a unique IMT \cite{first}. Nominally, there is one 3$d$ electron per Ti site in {\TiO}. The nearest neighbor Ti ions along the $c$-axis show dimerization in the crystal lattice and the two electrons on the Ti-Ti dimer form a singlet pair below the insulator-metal transition temperature {\Tim} ($\sim 450$ K). As temperature increases, the dimerization becomes weaker and the metallic state emerges. The low-temperature insulating state cannot be reproduced by the LDA calculations without the explicit consideration of the on-site Coulomb correlation \cite{LDA}. There are two possible scenarios to explain the IMT. One is the band crossing scenario \cite{good,zeig,zeig2}; the IMT is viewed as the transition from the dimerized insulator to the semimetal. If the singlet state is destabilized, the conducting {\egp} state may cross the Fermi energy. The other is the Mott transition \cite{sigma}. In this scenario, the formation of the singlet state is compatible with the large on-site Coulomb repulsion, and the electronic structure is expected to be totally reconstructed on a large energy scale in the course of the Mott transition as observed in the optical conductivity spectra for a similar dimer system {\VOO} \cite{VO2opt2}. It is known that the metallic state remains down to the lowest temperature when Ti ions are partially replaced by V ions \cite{resV}. In order to elucidate the origin of the IMTs in {\TiO}, we have investigated the thermally and doping induced IMTs in {\TiVO} by optical and transport measurements. We observe the pseudogap feature in optical spectra even in the metallic state. Drude weight as well as the carrier number obtained by the Hall measurement at low temperature is proportional to the V concentration. These features are consistent with the picture of the doped band insulator, while the strong electron correlation on Ti site is essential for the robust singlet formation.  

Single crystals of {\TiVO} with $x=0$, 0.006, 0.02, and 0.06 were prepared by the floating-zone method in Ar ($93 \%$) and H$_2$ ($7 \%$) atmosphere. The longitudinal and Hall resistivities were measured with PPMS (Quantum Design). Polarized reflectivity in the temperature range of $10-600$ K were measured between 0.01 eV and 5 eV. The measurements above room temperature were performed in the Ar/H$_2$ or vacuum atmosphere. Reflectivity spectra above 5 eV (up to 40 eV) were measured at room temperature with use of the synchrotron radiation at UV-SOR, Institute of Molecular Science, Okazaki. For the Kramers-Kronig analysis to deduce the optical conductivity, the spectrum above 40 eV was extrapolated by $\omega ^{-4}$ function, while below 0.01 eV the Hagen-Rubens relation and the constant reflectivity were assumed for metallic and insulating samples, respectively. Variation of these extrapolation procedures was confirmed to cause negligible difference in the conductivity spectrum.

We show the temperature and doping dependences of the resistivity in Fig. \ref{fig1}(c). The resistivity for {\TiO} shows a steep increase around {\Tim} \cite{first}. The resistivity for the current parallel to the $c$-axis ({\bf I} $\parallel c$) is about three times as high as that for {\bf I} $\perp c$ \cite{seebeck_resA2_Eg}. For $x=0.006$, the resistivity is much decreased but shows an insulating behavior. Above $x=0.02$, the resistivity shows a metallic behavior, and for $x=0.06$ the increase of the resistivity around {\Tim} is hardly observed. The anisotropy is observed even in the metallic state up to $x=0.06$. The measurements of the Hall resistivity \cite{hall_Eg} and the Seebeck coefficient \cite{seebeck_resA2_Eg} indicate that V ion is nominally divalent in {\TiVO} and thus the holes are introduced into the valence band (see also Fig. \ref{fig3}(d)). 

\begin{figure}
\begin{center}
\includegraphics*[width=8.6cm]{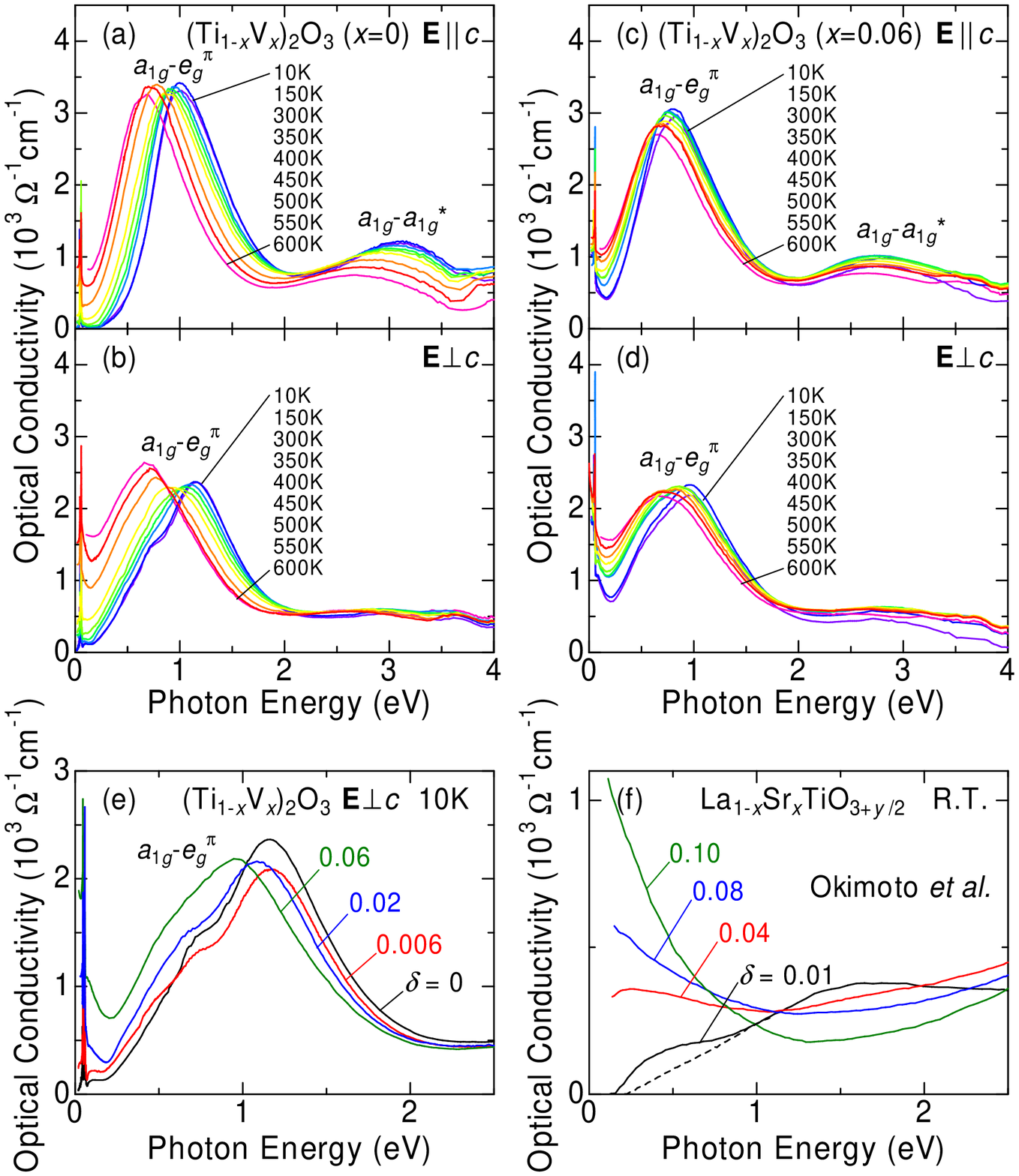}
\caption{(color online). Optical conductivity spectra at various temperatures in {\TiVO} with $x=0$ for (a) {\EparaC} and (b) {\EperpC}, and with $x=0.06$ for (c) {\EparaC} and (d) {\EperpC}. Hole-doping dependence of the optical conductivity spectra in (e) {\TiVO} and (f) {\LSTO} \cite{LSTOopt}. $\delta =x$ (or $\delta =x+y$) is nominal hole concentration, i.e., $\delta =1-n$ where $n$ is the number of $d$ electrons per Ti site. A dashed straight line in (f) indicates the hypothetical energy gap feature for $\delta=0$ .}
\label{fig2}
\end{center}
\end{figure}
In Figs. \ref{fig2}(a)-(d), we show the optical conductivity spectra at various temperatures in {\TiVO} with $x=0$ and 0.06 for {\EparaC} and {\EperpC}. For {\TiO}, the {\EparaC} spectrum shows a clear gap structure of about 0.2eV at the ground state (10 K). (Sharp spikes below 0.1 eV are due to the optical phonon modes \cite{fir}.) The proposed electronic structure of {\TiO} \cite{good} is shown in Fig. \ref{fig1}(b). The {\ttwog} levels in cubic symmetry split into the {\aoneg} and {\egp} levels due to the trigonal ligand field. Furthermore, the {\aoneg} band splits into the bonding {\aoneg} and the antibonding {\aonegs} bands mainly because of the hybridization within the Ti-Ti dimer. Two peaks around 1 eV and 3 eV are assigned to the {\aoneg}-{\egp} and {\aoneg}-{\aonegs} interband transitions, respectively \cite{nir}. The {\aoneg}-{\aonegs} dipole transition in the dimer is prohibited for {\EperpC}. Because of the anisotropic extension of the {\aoneg} orbital along the $c$-axis, the {\aoneg}-{\egp} transition appears stronger for {\EparaC} than for {\EperpC}. With increasing temperature, these peaks markedly shift to lower energy by about 0.4 eV. Nevertheless, the clear band gap feature remains to be observed in the whole temperature up to 600 K. For {\EperpC}, a similar gap structure as well as the conspicuous low-energy shift of the interband transition is also observed. This result is incompatible with the theoretical picture for the IMT of {\TiO} based on the Mott-Hubbard scheme \cite{sigma}. For $x=0.06$, a Drude peak is observed in the low-energy region, while the peak originating from the {\aoneg}-{\egp} transition is robust as in the high-temperature region of {\TiO}. In Fig. \ref{fig2}(e), we show the doping dependence of the ground-state optical conductivity spectra for {\EperpC} in {\TiVO} at 10K. With V-doping, the peak energy of the {\aoneg}-{\egp} transition decreases, while the Drude response evolves below 0.2 eV. Nevertheless, the Drude weight remains small compared with the spectral weight of the interband transitions within the {\ttwog} manifold. 

\begin{figure}
\begin{center}
\includegraphics*[width=8.6cm]{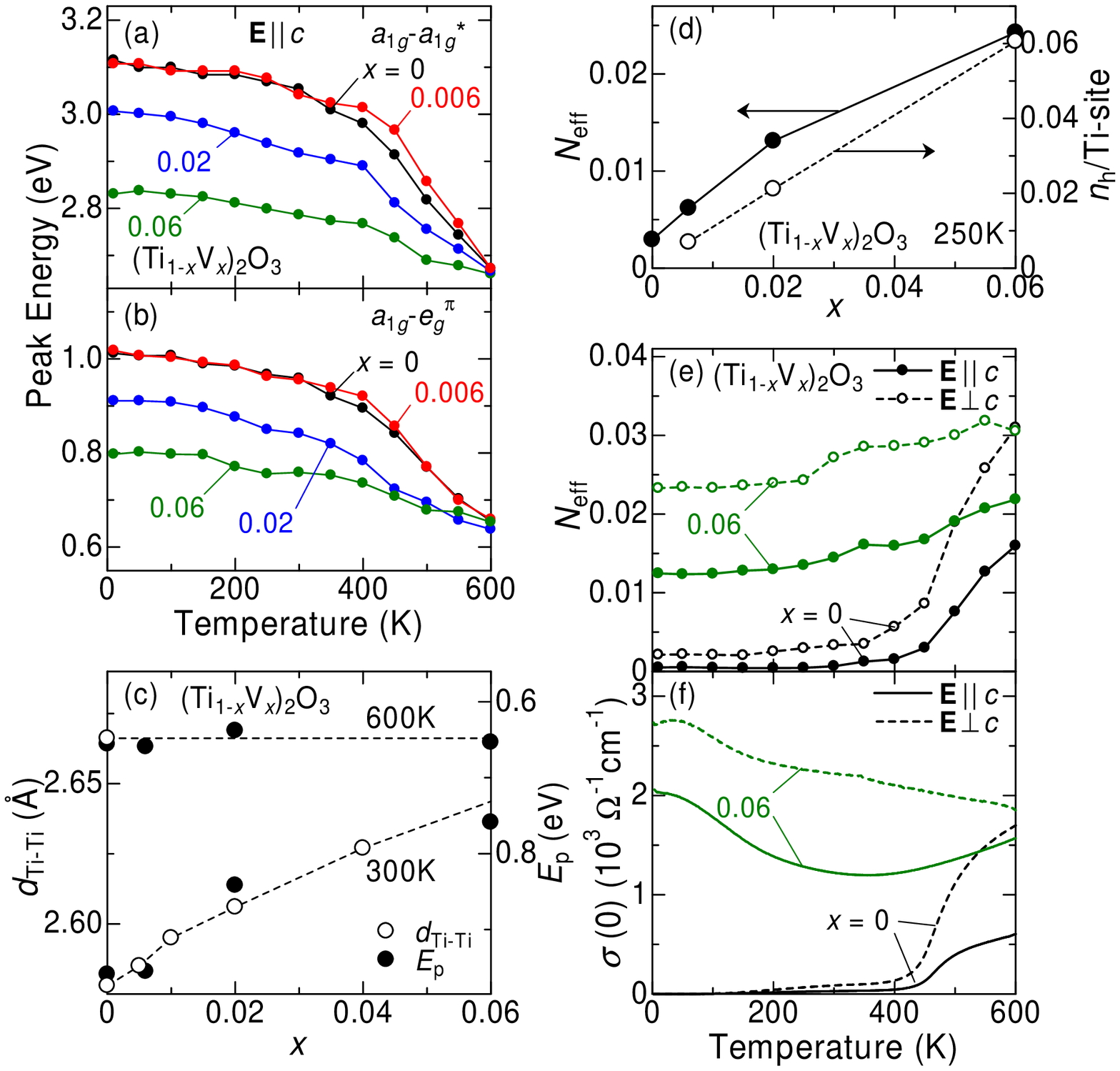}
\caption{(color online). Temperature variation of the peak energies for the interband transitions (a) {\aoneg}-{\aonegs} and (b) {\aoneg}-{\egp} for {\EparaC} in {\TiVO}. (c) Temperature and doping variation of the Ti-Ti dimer bond length $d_{\mathrm{Ti-Ti}}$ (open circles) reproduced from the structural data in literature \cite{crysT,crysV1} in comparison with the {\aoneg}-{\egp} ({\EparaC}) peak energy $E_{\mathrm{p}}$ (closed circles). (d) $x$ dependence of the hole type carrier density $n_{\mathrm{h}}$ per Ti site and the effective number of the electrons {\Neff} at 0.2 eV for {\EperpC} as a measure of the Drude weight (see text). Temperature dependence of (f) the {\Neff} at 0.2 eV and (g) the conductivity {\dc} in {\TiVO} ($x=0$ and 0.06).}
\label{fig3}
\end{center}
\end{figure}
In Figs. \ref{fig3}(a) and (b), the peak energies for the {\aoneg}-{\aonegs} and {\aoneg}-{\egp} interband transitions for {\EparaC} in {\TiVO} are plotted as a function of temperature, respectively. The shift of the peaks is clearly related with the change of the {\aoneg}, {\egp}, and {\aonegs} band positions. Both peak energies for {\EparaC} in {\TiO} are nearly temperature independent at low temperatures but show rapid decreases around {\Tim}. The amount of the shift is comparable to, or even larger than, the band gap $E_{\mathrm g}$ ($\sim $ 0.2 eV) in {\TiO} below room temperature. As $x$ increases, the peak energy at low temperatures also reduces. However, the peak position around 600 K is almost independent of $x$, indicating that the conducting state above {\Tim} is common in nature irrespective of the V-doping level. In Figs. \ref{fig3}(c), we plot the peak energy $E_{\mathrm{p}}$ of the {\aoneg}-{\egp} transition for {\EparaC} at 300 and 600 K in comparison with $x$ dependence of the Ti-Ti dimer bond length $d_{\mathrm{Ti-Ti}}$ estimated from the $c/a$ ratio \cite{crysT,crysV1}. Notably, the temperature and doping variation of the interband transition energy is parallel to that of the $d_{\mathrm{Ti-Ti}}$. The V-doping induces the hole type carriers at a rate of $x / {\mathrm{Ti}}$-site, as shown in Fig. \ref{fig3}(d). Therefore, the hole accumulation in the {\aoneg} valence band appears to relax the Ti-Ti dimer bond, leading to the reduction of the energy gap between the {\aoneg} and {\egp} bands. These results unambiguously show that the local singlet formation on the Ti-Ti dimer is responsible for the robust band gap but amenable to the thermal band-gap excitation or the hole-doping. 

\begin{figure}
\begin{center}
\includegraphics*[width=7cm]{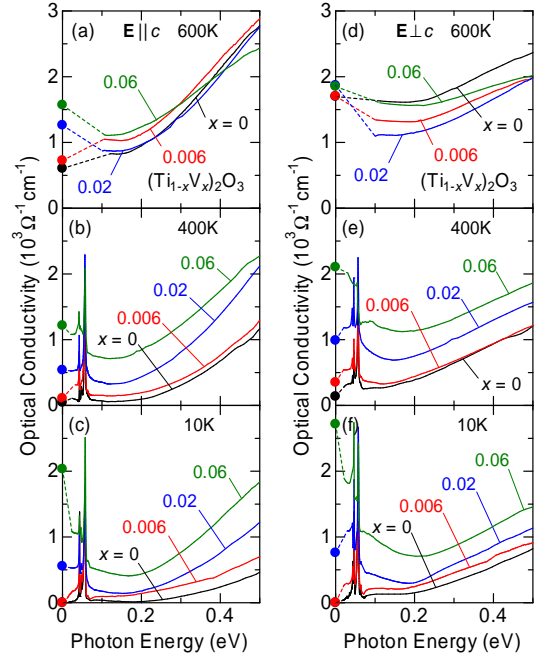}
\caption{(color online). Magnified view of the optical conductivity spectra below 0.5 eV in {\TiVO} at (a) 600 K, (b) 400 K, and (c) 10 K for the polarization {\EparaC} and (d)-(f) those for {\EperpC}. Closed circles represent the measured dc conductivity in each $x$. Dashed lines interpolating between the dc and optical conductivities are merely the guide to the eyes.}
\label{fig4}
\end{center}
\end{figure}
To see the carrier dynamics in more detail, we show in Fig. \ref{fig4} the low-energy part of the optical conductivity spectra in {\TiVO} at various temperatures below and above {\Tim}. At 600 K, the low-energy optical conductivity shows rather $\omega$-flat and less $x$-dependent features both for {\EparaC} and {\EperpC}, indicating the incoherent carrier dynamics in the conducting state above {\Tim}. The spectral weight in the low-energy region decreases with decreasing temperature for any $x$. Above $x=0.02$, however, the Drude-like coherent response is observed to grow below 0.2 eV as the temperature decreases.

Figures \ref{fig3}(d) and (e) show the doping and temperature dependence of the effective number of electrons {\Neff} at 0.2 eV in {\TiVO}, respectively, as a measure of Drude weight. The {\Neff} is defined by the following relation,
\begin{equation}
\label{neff}
\begin{split}
N_{\mathrm{eff}}=\frac{2m_{\mathrm{0}}V}{\pi e^2}\int_0^{\omega_{\mathrm{c}}}\sigma (\omega) d\omega.
\end{split}
\end{equation}
Here $m_{\mathrm{0}}$, $V$, and $\omega_{\mathrm{c}}$ ($=0.2$ eV) are the free-electron mass, the cell volume containing one Ti atom, and the cut off energy adopted as the {\aoneg}-{\egp} gap energy, respectively. (The contribution of the optical phonon was subtracted in Figs \ref{fig3}(d) and (e).) As seen in Fig. \ref{fig3}(d), the {\Neff} at 250K is nearly proportional to the doping level $x$, as well as the hole density $n_\mathrm{h}$. Thus, the metallic conductivity in the V-doped crystal is caused by the doped holes accumulated in the {\aoneg} valence band, whose concentration is proportional to $x$ as in the doped semiconductor; the conventional Drude formula suggests that $m^{\mathrm{*}}$ (effective mass) $\sim 4m_{\mathrm{0}}$ at low temperatures. The {\Neff} for {\TiO} shows a sudden increase around {\Tim} (Fig. \ref{fig3}(e)) and its temperature dependence roughly agrees with that of the dc conductivity {\dc} for both {\EparaC} and {\EperpC} (Fig. \ref{fig3}(f)). For $x=0.06$, the {\dc} increases with decreasing temperature below {\Tim}, while the {\Neff} decreases monotonically. This can be explained by the temperature dependence of the scattering rate (or mobility) of the carriers in the metallic state. The {\Neff} values at 600 K suggest the effective carrier density as low as $0.06-0.09 / {\mathrm{Ti}}$ in the high-temperature conducting state.

The comparison of these results with those in Mott transition system is meaningful. We show in Fig. \ref{fig2}(f) the doping variation of the optical conductivity spectra in the typical Mott transition system {\LSTO} (LSTO) reported by Okimoto {\etal} \cite{LSTOopt}. In LSTO, the large amount of the spectral weight is transferred from the high-energy Mott gap excitation to the low-energy Drude part in the course of the doping ($x+y / {\mathrm{Ti}}$-site) induced IMT. This behavior is quite contrastive with that in {\TiVO} shown in Fig. \ref{fig2}(e). The coexistence of the pseudo gap feature and the Drude peak with small spectral weight in {\TiVO} is consistent with the picture of the doped band insulator. Nevertheless, the doping induced low-energy shift of the whole interband transition bands (see Figs. \ref{fig3}(a) and (b)) is appreciable. Thus, the band gap arising from {\aoneg}-electron singlet formation in the bond dimer is critically affected by the hole doping level or the thermal gap excitation above {\Tim}. The thermally induced IMT in {\TiO} can be interpreted with band-crossing picture via the weakened singlet bond by the thermal gap excitation. The recent LDA+DMFT calculation suggests that the effective band width is reduced by the electron correlation, which stabilizes the dimerized state as well \cite{DMFT}.   

In summary, we have investigated the optical and transport properties for {\TiVO} in the wide temperature range of $2-600$ K. We identify that in the optical conductivity spectra the peaks due to the interband transitions remain to be observed in the thermally and V-doping induced metallic state. The variation of the gap energy in the course of the IMT well scales with that of the Ti-Ti bond length reflecting the singlet-bond strength. The Drude component is small compared with the weight of the interband transition bands and proportional to $x$, as well as the carrier number obtained by the Hall coefficient, whereas the magnitude of the band gap is sensitively affected by the thermal gap excitation and the hole doping. These observations are contrasted with the case of the canonical Mott transition in correlated electron systems, which accompanies the large spectral weight transfer from the gap excitation to the Drude weight. 

This work was partly supported by Grants-In-Aid for Scientific Research (Grant No. 15104006, 16076205, 20340086) from the MEXT of Japan.


\begin{thebibliography}{100}
\bibitem{MIT} M. Imada,  A. Fujimori, and Y. Tokura, Rev. Mod. Phys. {\bf 70,} 1039 (1998).
\bibitem{V2O3opt} M. J. Rozenberg {\etal}, Phys. Rev. Lett. {\bf 75,} 105 (1995).
\bibitem{LSTOopt} Y. Okimoto {\etal}, Phys. Rev. B {\bf 51,} 9581 (1995).
\bibitem{HTSC} T. Timusk and B. Statt, Rep. Prog. Phys. {\bf 62,} 61 (1999).
\bibitem{RVB} P. A. Lee, N. Nagaosa, and X.-G. Wen, Rev. Mod. Phys. {\bf 78}, 17 (2006).
\bibitem{first} F. J. Morin, Phys. Rev. Lett. {\bf 3,} 34 (1959).
\bibitem{LDA} L. F. Mattheiss, J. Phys.: Condens. Matter {\bf 8,} 5987 (1996).
\bibitem{good} L. L. Van Zandt, J. M. Honig, J. B. Goodenough, J. Appl. Phys. {\bf 39,} 594 (1968).
\bibitem{zeig} H. J. Zeiger, T. A. Kaplan, and P. M. Raccah, Phys. Rev. Lett. {\bf 26,} 1328 (1971).
\bibitem{zeig2} H. J. Zeiger, Phys. Rev. B {\bf 11,} 5132 (1975).
\bibitem{sigma} A. Tanaka, J. Phys. Soc. Jpn. {\bf 73,} 152 (2004).
\bibitem{VO2opt2} M. M. Qazilbash {\etal}, Phys. Rev. B {\bf 74,} 205118 (2006).
\bibitem{resV} G. V. Chandrashekhar {\etal}, Mat. Res. Bull. {\bf 5,} 999 (1970).
\bibitem{seebeck_resA2_Eg} S. H. Shin {\etal}, Phys. Rev. B {\bf 8,} 1364 (1973).
\bibitem{hall_Eg} J. M. Honig and T. B. Reed, Phys. Rev. {\bf 174,} 1020 (1968).
\bibitem{fir} G. Lucovsky, R. J. Sladek, and J. W. Allen, Phys. Rev. B {\bf 16,} 5452 (1977).
\bibitem{nir} S. S. M. Lu, F. H. Pollak, and P. M. Raccah, Phys. Rev. B {\bf 17,} 1970 (1978).
\bibitem{crysT} C. N. R. Rao, R. E. Loehman, and J. M. Honig, Phys. Lett. {\bf 27A,} 271 (1968).
\bibitem{crysV1} C. E. Rice and W. R. Robinson, J. Solid State Chem. {\bf 21,} 145 (1977).
\bibitem{DMFT} A. I. Poteryaev, A. I. Lichtenstein, and G. Kotliar, Phys. Rev. Lett. {\bf 93,} 086401 (2004).
\end{thebibliography}
\end{document}